\documentclass[epj]{svjour}
\usepackage{epsf}
\begin{document}

\title{Persistent currents for Coulomb interacting electrons on 2d 
       disordered lattices}
\subtitle{Sign and interaction dependence in the Wigner crystal regime}
\titlerunning{Persistent currents in the Wigner crystal regime}

\author{Franck Selva\inst{1} \and Dietmar Weinmann\inst{2,3}
\thanks{\email{weinmann@ipcms.u-strasbg.fr}}}
\authorrunning{Franck Selva and Dietmar Weinmann}

\institute{CEA, Service de Physique de l'Etat Condens\'e,
           Centre d'Etudes de Saclay, 91191 Gif-sur-Yvette, France \and
          Institut de Physique et Chimie des Mat\'eriaux de Strasbourg, 
           UMR 7504, CNRS-ULP, 23 rue du Loess, 67037 Strasbourg, France \and 
          Institut f\"ur Physik, Universit\"at Augsburg, 86135 
          Augsburg, Germany}

\date{Received 14 March 2000}

\abstract{
A Wigner crystal structure of the electronic ground state is induced by 
strong Coulomb interactions at low temperature in clean or disordered 
two-dimensional (2d) samples.
For fermions on a mesoscopic disordered 2d lattice, 
being closed to a torus, we study the persistent current in the 
regime of strong interaction at zero temperature. We perform a perturbation 
expansion starting from the Wigner crystal limit which yields power laws for 
the dependence of the persistent current on the interaction strength.  
The sign of the persistent current in the strong interaction limit
is independent of the disorder realization and strength. 
It depends only on the electro-statically determined configuration of
the particles in the Wigner crystal. 
} 

\maketitle

\section{Introduction} 

 The interplay of disorder and interaction in mesoscopic samples has 
attracted considerable interest in recent years \cite{Moriond}. A prominent 
experimental finding in this field is the insulator-metal transition 
in 2d systems of high mobility \cite{kravchenko}, 
occurring when the carrier density is increased. This transition can
neither be explained by disorder effects alone nor by interaction
effects in clean samples and is the subject of still increasing
experimental and theoretical activities (for a review see
\cite{mit-review} and references therein). 

 One possible mechanism invokes both, interaction and disorder
and associates the insulator--metal transition with the melting of
a pinned Wigner crystal when the carrier density increases and therewith
$r_{\rm s}$ (the interaction energy in units of the kinetic energy) decreases, 
allowing for metallic behavior at intermediate $r_{\rm s}$. Only at a higher 
density, when $r_{\rm s}$ is small, the interaction becomes negligible and 
the now dominating disorder-induced Anderson localization leads back to 
insulating behavior. Such a scenario is supported by experiments 
\cite{hamilton} and numerical investigations of few interacting particles in 
disordered lattice models \cite{benenti,benenti_mor,waintal_ann}.

 Another important experimental result in mesoscopic physics whose 
explanation necessitates to invoke disorder and interactions simultaneously 
is the value of the persistent current in diffusive rings 
\cite{pc_exp}. These persistent currents are much larger than the theoretical 
prediction for non-interacting electrons in disordered rings \cite{pc_nonint}. 
While the electron-electron interaction seems to play an essential
role, the disorder in the sample is also important: 
Interactions cannot affect the persistent current in clean
rotationally invariant 1d rings \cite{shastry,axel}, and the non-interacting
result is consistent with the experimental one for a clean semiconductor
ring in the ballistic regime \cite{mailly}. 

This has generated a large theoretical activity, dealing
with the combined effect of interactions and disorder on the
enhancement of persistent currents in mesoscopic rings 
(for an overview see e.g. \cite{Moriond,Imry_book,eckern} and
references therein).
Even though different theoretical approaches suggest an increase of the 
persistent current in disordered samples due to repulsive Coulomb 
interactions, a quantitative understanding of the experiments is 
still lacking. 

Within {\em continuous} models of fermions in 1d with disorder, 
repulsive interactions are found to enhance the persistent 
currents, without \cite{axel} and with \cite{cohen} spin. 
On moderately disordered {\em lattices}, repulsive 
interactions are however found to decrease the persistent current for 
spinless fermions in 1d \cite{berko3,bouzerar,kato}. 
 It was concluded from analytical considerations \cite{gs}, using 
renormalization group arguments, that 1d lattice models exhibit an
interaction-induced enhancement of the persistent current only when 
the spin degree of freedom is taken into account. Nevertheless, a
numerical study of small systems \cite{abraham} revealed that
repulsive interactions can slightly enhance the averaged persistent current
even for spinless fermions on a 1d lattice, provided the disorder is
very strong.

 A recent numerical investigation treating the persistent current for 
strongly disordered {\em individual} 1d chains \cite{schmitteckert,WPSJ} 
leads to the conclusion that, while Anderson localization is dominating the 
non-interacting case, the persistent current can be {\em strongly enhanced} by 
repulsive local interactions at sample dependent intermediate values of the 
interaction strength. This delocalization is accompanied by a reorganization 
of the ground state structure. At half filling, strong interactions induce a 
regular charge density in the Mott-insulator regime and decrease the 
persistent current. 

This suppression of the persistent current in the limit of strong
repulsive interactions is not limited to 1d models of spinless
fermions at half-filling with local interactions. It occurs with
long-range interactions \cite{berko3} at arbitrary filling and also 
in 2d lattices \cite{benenti}, when strong interaction (large $r_{\rm s}$)
leads to the formation of a Wigner crystal pinned by the disorder in
the ground state.

In the case of the Hubbard model at half
filling \cite{stafford} and for spinless fermions in 1d chains without
\cite{tsiper1} and with disorder \cite{WSJP}, this suppression of the 
persistent current at strong interaction has been understood
quantitatively from a perturbation theory starting at
the Mott insulator limit. In contrast, numerical Hartree-Fock
approaches which allow to treat larger systems than the exact
diagonalization used in Refs.~\cite{benenti,berko3}
and which have been used for weakly interacting particles in 1d and 2d 
\cite{bouzerar2}, are, at least in 1d, not able to quantitatively describe the
persistent current in the limit of strong 
interaction \cite{piechon}.

The sign of the persistent current in the case of spinless particles 
in strictly 1d rings is independent on the disorder realization and the 
interaction strength. According to a well-known theorem by Leggett 
\cite{leggett}, the sign is given solely by the parity of the particle
number $N$ (paramagnetic for $N$ even and diamagnetic for $N$ odd).
This general rule is confirmed by an explicit calculation for a
Luttinger liquid ring without disorder \cite{Loss}.

On the other hand, the sign of the persistent current for particles
with spin in disordered 1d rings or spinless fermions in disordered 
2d systems differs from sample to sample and no general sign rule exists. 
Only in some special cases, like non-interacting particles with spin
in clean 1d rings \cite{loss-gold}, and for some interacting
situations using the Hubbard model \cite{fye,yu}, the
sign of the persistent current has been determined.  
The situation is less clear in the presence of long-range 
interactions and disorder we address in this work.
For electrons with spin in 1d rings having a particular disorder
consisting of only one barrier, a tendency towards diamagnetic
responses, independent of the particle number, was found at 
strong repulsive interaction \cite{haeusler}.

For spinless fermions in strongly disordered 2d lattice models, it has
been noticed in numerical studies that the sign of the
persistent current becomes realization-independent in the limit of
strong interaction \cite{benenti,benenti_mor,berko1,berko2}. Detailed 
studies of the local current show that the suppression of its transverse
component by the interactions is much stronger than the decrease of
the longitudinal current.
On such a lattice, closed to a torus, the structure of the ground state
at strong interaction is a Wigner crystal pinned by the disorder 
\cite{benenti,benenti_mor,berko2}. While the system exhibits Anderson 
localization at weak interaction, the regime of intermediate interaction
shows indications of a new type of correlated metal \cite{waintal}. 

For spinless fermions in 2d lattice models {\em without disorder}, the
amplitude of the persistent current has recently been studied
analytically and numerically at strong interaction \cite{tsiper2}. 
When $r_s$ is large, the hopping matrix elements between neighboring 
lattice sites being much smaller than the interaction strength, the 
behavior can be understood from a perturbation theory in terms of the
hopping matrix elements. 

In this paper we report a study of the persistent current in 
{\em disordered} 2d lattice models at {\em very strong} Coulomb 
interaction, using a perturbation theory expansion around the pinned 
Wigner crystal. 
While the understanding of this regime cannot directly explain
the high persistent currents observed experimentally in diffusive
metal rings, it may be relevant for the insulating side at low carrier
density of the insulator-metal transition in 2d. We show how  
the sign of the persistent current at strong interaction follows 
systematically from the structure of the Wigner crystal and find
simple rules for this sign. For the absolute values, power laws
similar to the ones found in \cite{tsiper2} are obtained also in the 
disordered case. We shall show how the presence of disorder and 
spin can influence the prefactors and the exponents of these power laws.

In the following section we introduce the model for interacting fermions on 
a disordered lattice and the quantities used to characterize its properties.
The perturbation theory is developed in section \ref{sec:theory} and 
applied in section \ref{sec:pc} to the persistent current in longitudinal and 
transverse direction before we conclude the paper.

\section{Model}
\label{sec:model}

\subsection{Hamiltonian}

We consider $N$ fermions on a disordered square lattice with Coulomb 
interaction. The corresponding Hamiltonian reads
\begin{equation}\label{hamiltonian}
H=H_{\rm K}+H_{\rm D}+H_{\rm I}\, .
\end{equation} 
The kinetic energy term is
\begin{equation}
H_{\rm K}= -t \sum_{<i,i'>}\sum_\sigma 
      c^{+}_{i,\sigma}c^{\phantom{+}}_{i',\sigma}
\end{equation}
with the hopping matrix element $t=1$ setting the energy scale. 
We concentrate on rectangular 2d lattice structures with $L_x \times L_y$ 
sites $i$. The fermionic on-site operators 
$c_{i,\sigma}\equiv c_{(x_i,y_i),\sigma}$ destroy
a particle with spin $\sigma$ located at $\vec{r}_i=(x_i,y_i)$, where the 
position coordinates $x_i\in \{ 1,2,\dots, L_x\} $ and 
$y_i\in \{ 1,2,\dots, L_y\} $ are measured in units of the lattice spacing $a$.

The disordered potential contribution is
\begin{equation}
H_{\rm D}= W \sum_{i,\sigma} v_i \hat{n}_{i,\sigma} \, ,
\end{equation}
where $W$ is the disorder strength, with the independent random variables 
$v_i$, drawn from a box distribution within the interval $[-1/2;+1/2]$.
The occupation number operators are as usual given by 
$\hat{n}_{i,\sigma}= c^{+}_{i,\sigma}c^{\phantom{+}}_{i,\sigma}$. 
The Coulomb interaction is described by the term
\begin{equation}
H_{\rm I}= \frac{U}{2} \sum_{i,i'\atop i\neq i'} \sum_{\sigma,\sigma'}
\frac{\hat{n}_{i,\sigma} \hat{n}_{i',\sigma'}}{|\vec{r}_i-\vec{r}_{i'}|}
+ U \sum_{i}\frac{\hat{n}_{i,\uparrow}\hat{n}_{i,\downarrow}}{d}\, ,
\end{equation}
the interaction strength being parametrized by $U$. 
The sum in $H_{\rm K}$ runs over all pairs of sites which 
are next neighbors on the lattice $<i,i'>$, while the interaction term 
is composed of a sum over all pairs of different sites.
With these definitions, one finds $r_{\rm s}=\frac{U}{2t\sqrt{\pi \nu}}$,
with the electronic density or filling factor $\nu=N/L_xL_y=1/b^2$, $b$
being the average distance between particles.
An additional term takes into account double occupancy of a site by two 
particles of different spin, $d \, (<a)$ being a measure for the size of the 
on-site orbitals.

In order to study the persistent current, we close the 2d lattice first to 
a cylinder by imposing generalized periodic boundary conditions 
\begin{equation}
c_{(L_x,y);\sigma}=\exp({\rm i}\phi_x) c_{(0,y);\sigma}
\end{equation}
in $x$-direction. For $\phi_x=0$, this is equivalent to usual periodic
boundary conditions. Finite $\phi_x$ accounts for a magnetic flux 
$\Phi=\phi_x \Phi_0/2\pi$ threading the ring, $\Phi_0$ being the flux quantum.
We choose the units such that $\Phi_0/2\pi=1$.
In order to reduce finite size effects, we use periodic boundary 
conditions in $y$-direction, and thus obtain a torus 
topology with the fluxes $(\phi_x,\phi_y)=(\phi_x,0)$. In section 
(\ref{sec:transverse}) we will use the dependence on the transverse
flux $\phi_y$ to study also the transverse current.

\subsection{Persistent current}

The magnetic flux threading the ring can drive a persistent current through 
the system. At zero temperature, it is given by
\begin{equation}
I(\phi_x)=-\left. \frac{\partial E_0}{\partial \phi}\right|_{\phi=\phi_x}\, ,
\end{equation}
where $E_0$ is the many-particle ground state energy. Thus, the persistent 
current at $T=0$ is a measure of the dependence of the ground state energy on
the magnetic flux. Since the latter can be expressed in the form of a boundary 
condition, it is at the same time a measure of the ground state sensitivity to
the boundary conditions and can be related to the conductance of the sample
\cite{kohn}. 

\section{Theoretical approach}\label{sec:theory}

\subsection{Wigner crystal at strong interaction}

In the non-interacting limit, disorder leads to Anderson localization of the 
one-particle states and the problem can be treated by a perturbative expansion 
around the on-site localized states in terms of the hopping matrix elements 
$t$ \cite{bouch89}. 
Hopping to distant sites costs disorder energy of the order of $W$ such
that a series expansion in $t/W$ results. 

In the many-body case, strong repulsive interaction $U$ leads to Wigner 
crystallization of the ground state with on-site localized charges in the 
electro-statically most favorable position. One can use a similar perturbative 
formalism in terms of the hopping $t$, but now, the essential cost in energy  
caused by deplacing one of the many particles is given by the increase
of the interaction energy such that one obtains a systematic expansion
in terms of $t/U$. 

We decompose the Hamiltonian (\ref{hamiltonian}) as
\begin{equation}
H=H_0+H_{\rm K}
\end{equation} 
with an unperturbed part containing disorder and interaction
\begin{equation}
H_0=H_{\rm D}+H_{\rm I} \, ,
\end{equation}
and the perturbation given by the hopping terms of $H_{\rm K}$.

$H_0$ is composed of terms containing only occupation number operators 
in the one-particle on-site basis. Therefore, its $N$-particle eigenstates
$|\psi_\alpha \rangle$ are Slater determinants built from $N$ 
different one-particle functions and are completely characterized by the 
occupation numbers $n_{i,\sigma}(\alpha)\in \{0,1\}$ of the one-particle 
states on site $i$ with spin $\sigma$, fulfilling the condition 
$N=\sum_{i,\sigma} n_{i,\sigma}(\alpha)$. 
Therewith, the many-body eigenstates of $H_0$ can be written in the form  
\begin{equation}\label{eq:states}
|\psi_\alpha \rangle =\left(\prod_{i,\sigma} 
(c^{+}_{i,\sigma})^{n_{i,\sigma}(\alpha)}\right) |0\rangle
\end{equation}
($|0\rangle $ is the vacuum state), and the corresponding eigenenergies are 
given by $E_\alpha=E^{\rm D}_\alpha+E^{\rm I}_\alpha$ with
\begin{eqnarray}
E^{\rm D}_\alpha &=& W\sum_{i,\sigma} v_{i} n_{i,\sigma}(\alpha) \quad\quad 
\mbox{and}\\  
\label{eq:int_energy}
E^{\rm I}_\alpha &=&\frac{U}{2} \sum_{i,i'\atop i\neq i'}\sum_{\sigma,\sigma'} 
\frac{n_{i,\sigma}(\alpha) n_{i',\sigma'}(\alpha)}{|\vec{r}_i-\vec{r}_{i'}|}
+ U \sum_{i}\frac{n_{i,\uparrow}n_{i,\downarrow}}{d}\, .
\end{eqnarray}
The ground state $|\psi_0\rangle$ of this Coulomb glass problem is given 
by purely classical considerations, minimizing disorder and interaction energy.
Its charge configuration depends in general on the specific disorder 
realization of the sample. 
At strong enough interaction, when the disorder effects are dominated by the 
interaction, the structure of $|\psi_0\rangle$ is the Wigner crystal 
of minimal interaction energy, independent of the disorder realization.
The rigid array of charges can be translated as a whole through the 
system without changing the interaction energy. Nevertheless, the 
contribution of the disordered potential to the energy depends on the 
realization and pins the Wigner crystal in a realization dependent position.
It is important to realize that the structure of the Wigner crystal is 
entirely given by the lattice geometry and the Coulomb interaction of the 
$N$ particles. 

In contrast to $H_0$, the perturbing part
$H_{\rm K}$ of the Hamiltonian depends on the magnetic flux through the ring 
since the latter appears in the boundary condition in $x$-direction and 
therefore influences some hopping matrix elements.
Writing all the hopping terms explicitly, one obtains 
\begin{eqnarray}
&&H_{\rm K}(\phi_x,\phi_y)= -t \sum_{\sigma} \left( 
\sum_{x=1}^{L_x} \sum_{y=2}^{L_y} 
c^+_{(x,y);\sigma} c^{\phantom{+}}_{(x,y-1);\sigma} \right. \nonumber \\
&& + \sum_{x=2}^{L_x} \sum_{y=1}^{L_y} 
c^+_{(x,y);\sigma} c^{\phantom{+}}_{(x-1,y);\sigma} 
 + e^{-{\rm i}\phi_y} \sum_{x=1}^{L_x} 
c^+_{(x,1);\sigma}c^{\phantom{+}}_{(x,L_y);\sigma} \nonumber \\
&&  \left.  + e^{-{\rm i}\phi_x} \sum_{y=1}^{L_y} 
c^+_{(1,y);\sigma}c^{\phantom{+}}_{(L_x,y);\sigma} + \mbox{H.C.}\right)\, . 
\end{eqnarray}

\subsection{Perturbation expansion for spinless fermions}

For electrons with spin, the unperturbed ground state is $2^N$-fold degenerate
since all the spin configurations yield the same energy 
when hopping is suppressed. Further degeneracies appear in the case of clean 
systems when $W=0$ allows for translational symmetry. 

For simplicity we first 
treat the case of completely spin polarized systems (all spins up, equivalent 
to spinless fermions) with disorder where the ground state of $H_0$ is
not degenerate and the expansion in $H_{\rm K}$ is straightforward. 

Using standard perturbation theory, the correction to the ground state
energy in $n^{\rm th}$ order is given by
\begin{eqnarray}\label{pert_n}
&&E_0^{(n)}=\sum_{\alpha_1,\alpha_2,\dots ,\alpha_{n-1}} \\
&&\frac{
\langle \psi_0 | H_{\rm K} | \psi_{\alpha_1} \rangle \langle \psi_{\alpha_1} | 
 H_{\rm K} | \psi_{\alpha_2} \rangle 
 \dots 
\langle \psi_{\alpha_{n-1}} | H_{\rm K} | \psi_0 \rangle }
{(E_0-E_{\alpha_1})(E_0-E_{\alpha_2})\dots (E_0-E_{\alpha_{n-1}})} \, , 
\nonumber
\end{eqnarray}  
with the sums running over all the eigenstates of $H_0$ except the 
ground state itself.

The numerator of the contributions to the sum of equation (\ref{pert_n}) 
contains matrix elements 
$\langle\psi_{\alpha_{k}} | H_{\rm K} |\psi_{\alpha_{k+1}}\rangle $ 
of the perturbing Hamiltonian. Since $H_{\rm K}$ consists only of one-particle 
hopping terms, non-zero matrix elements can arise only if the two states 
$|\psi_{\alpha_k}\rangle$ and $|\psi_{\alpha_{k+1}}\rangle$ differ by nothing 
else than a single hop of one of the particles. 
From (\ref{pert_n}) one sees that a finite contribution to the sum over 
different sequences of intermediate states $\alpha$ is obtained only if the
$n$ one-particle hops are such that the final configuration has an
overlap with the initial one, corresponding to the ground state. The $n$ sums 
over intermediate states $\alpha_k$ in equation (\ref{pert_n}) can then be 
rewritten as a sum over all the sequences 
$S=(\alpha_1,\alpha_2,\dots,\alpha_{n-1})$ which give a nonzero contribution. 
Denoting the numerator of the terms by ${\rm Num}(S)$ and the denominator
by ${\rm Den}(S)$, equation (\ref{pert_n}) takes the form  
\begin{equation}\label{pert_S}
E_0^{(n)}=\sum_{S} \frac{ {\rm Num}(S) }{ {\rm Den}(S) }\, .
\end{equation}  
We will now evaluate the numerator and the denominator separately. 
The numerator ${\rm Num}(S)$ can be calculated directly from the $n$ hopping 
matrix elements, thereby taking into account the flux dependent phase for hops 
crossing the boundary. Since we consider fermions, the 
corresponding operators anti-commute and their order in the products 
(\ref{eq:states}) defining the basis states $|\psi_\alpha\rangle$ is crucial
for the sign. When the starting point (the ground state configuration) is 
re-established after $n$ consecutive hopping processes, the order of the 
operators can be modified. Then, the sign of the permutation $P_S$ of the 
operators, caused by the sequence of one-particle hops $S$, must be 
incorporated in the result. Altogether, one obtains 
\begin{equation}\label{numerator}
{\rm Num}(S) = {\rm sign}(P_S)\, 
(-t)^n \exp\left[-{\rm i}\phi_x(h_{\rm f}-h_{\rm b})\right]\, .
\end{equation}
$h_{\rm f}$ and $h_{\rm b}$ denote the number of hoppings across the
boundary between sites $(L_x,y)$ and $(1,y)$ in forward and backward direction,
respectively. Therefore, only the corrections to the ground state energy 
due to sequences with $h_{\rm f}-h_{\rm b}\neq 0$ are flux dependent. 

Moving particles creates defects in the Wigner crystal. This increases the 
interaction energy (see equation \ref{eq:int_energy})
of the ground state $E^{\rm I}_{0}$ by the amount $U\epsilon_\alpha$, where 
\begin{equation}\label{epsilon}
\epsilon_\alpha=\frac{1}{U}\left(E^{\rm I}_{\alpha}-E^{\rm I}_{0}\right) 
\end{equation}
is non-negative and independent of $U$. 
It accounts for the difference of the inter-particle distances between the 
configurations of the state $|\psi_\alpha\rangle$ and the ground state 
$|\psi_0\rangle$. We assume always $U\gg W$, such that the difference in 
potential energy can be neglected in a first step except for the case 
$\epsilon_\alpha= 0$ which occurs for some special sample geometries and 
particle numbers. Corrections due to the disorder will be considered in Section
(\ref{sec:disorder}).

In the generic case when $\epsilon_\alpha > 0$ for all intermediate states, 
the energy differences in the denominator ${\rm Den}(S)$ are dominated by the 
difference in interaction energy. In the limit of strong interaction, one 
therefore gets 
\begin{equation}\label{denominator}
{\rm Den}(S) \approx (-U)^{n-1}\prod_{\alpha=1}^{n-1} \epsilon_\alpha \, .
\end{equation}

\section{Persistent current}
\label{sec:pc}

\subsection{Longitudinal current}

Since the longitudinal persistent current at zero temperature 
$I=-\partial E_0/\partial \phi_x$ is given by the flux dependence of the 
ground state energy, a perturbative treatment of the latter in terms of the 
hopping $t$ yields a systematic expansion of the persistent current.

Calculating the $n^{\rm th}$ order
correction $E_0^{(n)}(\phi_x)$ to the ground state energy, one gets 
the correction to the persistent current
\begin{equation}
I^{(n)}(\phi_x)=-\frac{\partial E_0^{(n)}}{\partial \phi_x}
\end{equation}
in $n^{\rm th}$ order in the perturbation $H_{\rm K}$.

\subsubsection{Relevant terms of the perturbation theory}

\begin{figure}[tb]
\centerline{\epsfxsize=2in\epsffile{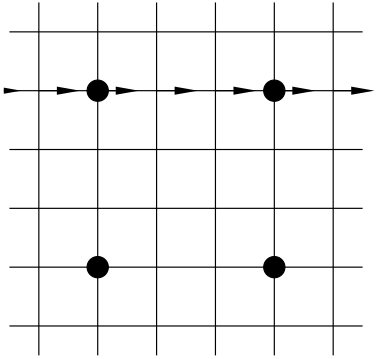}}
\caption[linehop]{\label{fig:linehop} 4 particles on $6\times 6$ sites. 
The Wigner crystal configuration has regular square structure. The arrows 
indicate $L_x=6$ hops from which a contribution in lowest order to 
the persistent current arises.}
\end{figure}

When the sequence $S$ of $n$ hopping elements is chosen such that each of the 
particles returns to its initial position without completing a tour around the 
ring, every particle which has crossed the boundary must necessarily cross it 
a second time in the opposite direction such that $h_{\rm b}=h_{\rm f}$. 
Therefore the flux dependence of these contributions disappears and they 
cannot influence the persistent current. The lowest order of the perturbation 
theory which yields a finite contribution $I^{(n)}$ is $n=L_x$, corresponding 
to the sequences in which one particle starting at $(x_0,y_0)$ crosses the 
boundary and returns to its original position after completion of its journey 
around the ring at constant $y=y_0$. If there are more than one particle in 
the line of the lattice with constant $y=y_0$, a contribution of the same 
order arises from sequences of hops which move each of the particles to the 
position of its neighbor (see Fig.\ \ref{fig:linehop}). Since any hopping in 
$y$-direction leads to an increase of the order of the contribution, 
the lowest order correction to the persistent current is given by the 
considered processes in which $y$ is kept constant and all of the hops are 
either in forward or in backward direction. 

We now address the dependence of the permutation $P_S$ corresponding to this 
kind of process on the number $N_y$ of particles in the line at constant $y$.
If $N_y=1$, the particle returns to its initial site, the final 
configuration is exactly equal to the initial one and ${\rm sign}(P_S)=+1$. 
In the general case of arbitrary $N_y$, the considered sequences of hops 
(see Fig.\ \ref{fig:linehop}) lead to a cyclic rotation of the order of the
$N_y$ particles and the sign of the corresponding permutation is 
${\rm sign}(P_S)=(-1)^{N_y-1}$. 

This yields the result for the lowest-order $\phi_x$-depen\-dence of the 
ground state energy 
\begin{equation}
E_0^{(L_x)}(\phi_x)\approx - \sum_{S}
\frac{t^{L_x} {\rm sign}(P_S)\exp\left[-{\rm i}\phi_x(h_{\rm f}-h_{\rm b})\right]}
{U^{L_x-1}\epsilon_{\alpha_1}\epsilon_{\alpha_2}\dots\epsilon_{\alpha_3} } \, .
\end{equation}
Each sequence considered in this sum contains either $L_x$ forward
hops with $h_{\rm f}=1$; $h_{\rm b}=0$ or it contains $L_x$ backward
hops with $h_{\rm f}=0$; $h_{\rm b}=1$. Each given backward sequence 
$S_{\rm b}$ can now be assigned to the forward sequence $S_{\rm f}$
with the reversed order of hops whose contribution differs only in the
sign of the flux-dependent phase-factor. 
One can express the result as a sum over the forward hopping sequences  
\begin{equation}\label{energy_lowest}
E_0^{(L_x)}(\phi_x)\approx -2 \frac{t^{L_x}}{U^{L_x-1}} \sum_{S_{\rm f}}
\frac{{\rm sign}(P_S)\, \cos\phi_x}{\epsilon_{\alpha_1}\epsilon_{\alpha_2}\dots
 \epsilon_{\alpha_{L_x-1}}} \, .
\end{equation}
Within the above perturbation theory, when processes corresponding to two 
loops around the ring (which are at least of order $2L_x$) are neglected 
as compared to the lowest order one-loop processes, the flux dependence of 
the ground state energy is harmonic and $2\pi$-periodic in $\phi_x$.

\subsubsection{Lowest order result for the persistent current}

From (\ref{energy_lowest}), one obtains the persistent current in $L_x$-th 
order perturbation theory
\begin{equation}\label{pc_lowest}
I^{(L_x)}(\phi_x) = \tilde{I}^{(L_x)} \sin\phi_x 
\end{equation}
with the flux--independent amplitude
\begin{equation}\label{amplitude_lowest}
\tilde{I}^{(L_x)} \approx -2 \frac{t^{L_x}}{U^{L_x-1}} \sum_{S_{\rm f}}
 \frac{{\rm sign}(P_S)}
{\epsilon_{\alpha_1}\epsilon_{\alpha_2}\dots \epsilon_{\alpha_{L_x-1}}}\, .
\end{equation}
This result contains several interesting features. 
First, the absolute value of the persistent current decays
proportionally to $t^{L_x}/U^{L_x-1}$, 
in the limit of strong interaction. In order to determine the constant 
prefactor of this power law, it is sufficient to figure out all possible 
processes which transform the ground state into itself, using $L_x$ forward 
hopping processes, and to calculate the corresponding $\epsilon_\alpha$ 
from (\ref{epsilon}).

Furthermore, the sign of 
the dominating contributions to the persistent current in the limit of strong 
interaction is given by $-{\rm sign}(P_S)$, which for spin-polarized electrons 
is given by $(-1)^{N_y}$ with the number $N_y$ of electrons in 
the line of the sample at constant $y$. This is consistent with the 
well-known theorem by Leggett for the sign of the persistent current of 
spinless fermions in 1d \cite{leggett} (positive $\tilde{I}$ or paramagnetic 
response for $N$ even and negative or diamagnetic for $N$ odd). 
Only if the unperturbed ground state (Wigner crystal) configuration of the 
particles is such that the particle numbers $N_y$ in different occupied 
lines $y$ have different parity, the prefactors of the corresponding terms 
have to be considered to determine the sign of the persistent current.
For $N_y$ particles in a line of length $L_x$, the number of hopping 
sequences $N_{\rm seq}(N_y)$ going from the ground state to itself 
is the number of terms contributing to the sum in equation
(\ref{amplitude_lowest}). 
Therewith, the result for the persistent current can be roughly
estimated to be
\begin{equation}
 \tilde{I}^{(L_x)}\propto \frac{t^{L_x}}{U^{L_x-1}}\sum_{y=1}^{L_y} 
  N_{\rm seq}(N_y) (-1)^{N_y}\, .
\end{equation}
In the
limit of low filling $N_y/L_x \rightarrow 0$, $N_{\rm seq}(N_y)$ is 
approximatively given by the number of possibilities $N_{\rm seq}(N_y)$ 
for $N_y$ particles to each make $L_x/N_y$ (here we assume for
simplicity that $L_x/N_y$ is an integer) forward jumps to reach the 
position of its neighbor, leading to the estimate
$N_{\rm seq}(N_y)\approx L_x!/[(L_x/N_y)!]^{N_y}$. 
At finite filling, one 
must consider that the neighbor particle must have left its starting site 
before the arriving particle can do its last hop. This correlation of the 
order of the hops of different particles reduces $N_{\rm seq}(N_y)$.
With increasing filling, $N_{\rm seq}(N_y)$ starts to exponentially 
increase with $N_y$ and continues to increase more slowly until 
$N_y=L_x/2$ (half filling). For larger filling it decreases, thereby 
obeying a symmetry with respect to half filling which is a consequence
of the symmetry between particles and holes. One can expect that the
contribution of the line $N_y=N_y^{\rm max}$ with the largest number
$N_{\rm seq}(N_y^{\rm max})$ of sequences (at low filling this is the
one with the maximum number of particles) may dominate over the 
contributions of the ones with fewer sequences. 
The sign of the persistent current is then likely to be
$(-1)^{N_y^{\rm max}}$.

\subsubsection{Examples}

As an example, we calculate explicitly the lowest order term in 
$\frac{1}{U}$ of the persistent current for 4 spinless fermions on a
few small $L_x \times L_y$ rectangular lattices, using the formula 
(\ref{amplitude_lowest}).  

\paragraph{$4\times 2$ sites}

\begin{figure}[tb]
\centerline{\epsfxsize=1.5in\epsffile{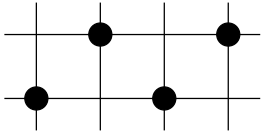}}
\caption[4x2]{\label{fig:4x2} Ground state configuration for 4 particles on a 
$4\times 2$ lattice.}
\end{figure}

We start with the simple case of a $4\times 2$ lattice. The electro-statically 
lowest energy configuration (the Wigner crystal) is shown in 
Fig.\ \ref{fig:4x2}.
The number of particles in the two lines at $y=1$ and $y=2$ of
the system is $N_y=2$, and the lowest order of the
perturbation theory which yields a contribution to the persistent current 
is $n=L_x=4$. From these two ingredients, we can immediately
determine the sign (${\rm sign} (P_S) = -1$ for $N_y$ even) and the
power law of the decrease of the persistent current at strong
interaction strength. 
The leading term of the amplitude (\ref{amplitude_lowest}) is given by
\begin{equation}\label{amp_4x2}
\tilde{I}^{(4)} \approx 2 \frac{t^{4}}{U^{3}} \sum_{S_{\rm f}}
 \frac{1}
{\epsilon_{\alpha_1}\epsilon_{\alpha_2}\dots \epsilon_{\alpha_{L_x-1}}}\, .
\end{equation}
This result is always positive and $\propto 1/U^3$. Therefore, the response 
of the system to the applied flux is always paramagnetic at strong interaction.

For this example, it is an easy exercise to figure out all hopping sequences 
which contribute in lowest order. For each line $y=1,2$, one finds 4 different 
sequences and can explicitly calculate the interaction energies and 
$\epsilon_\alpha$ of the intermediate states. This allows to evaluate the sum 
over all sequences in (\ref{amp_4x2}) with the result
\begin{equation}\label{I_4x2} 
\tilde{I}\rightarrow \tilde{I}^{(4)} \approx 853 \frac{t^{4}}{U^{3}}
\quad\quad \mbox{for} \quad\quad \frac{t}{U}\rightarrow 0 \, .
\end{equation}

In the case of 8 spinless fermions on $4\times 4$ sites, the same
analysis can be carried out except for the fact that now four lines 
$y=1,2,3,4$ must be considered, yielding
an additional factor of 2 in (\ref{I_4x2}). In particular, the
persistent current is always paramagnetic in the strong interaction
limit, as found numerically in Ref.\ \cite{berko2}.

\paragraph{$2\times 4$ sites}

The situation for 4 particles on $2\times 4$ sites is even simpler
since there are four lines $y=1,2,3,4$, each of them containing $N_y=1$ 
particle. Since $L_x=2$, the leading order of the perturbation theory at 
strong interaction is $n=2$, and the sequences of hops contain only
one intermediate state. The evaluation of (\ref{amplitude_lowest})
yields
\begin{equation}
\tilde{I}\rightarrow \tilde{I}^{(2)} \approx -15 \, \frac{t^{2}}{U}
\quad\quad \mbox{for} \quad\quad \frac{t}{U}\rightarrow 0 \, ,
\end{equation}  
and the persistent current is diamagnetic at strong interaction.

\paragraph{$6\times 6$ sites} 

4 particles on $6\times 6$ sites is the situation investigated numerically in 
Ref.\ \cite{benenti}, at strong disorder. As a function of the interaction 
strength, an increase of the average persistent current at intermediate 
strength and a decrease at strong interaction was found. An exponential 
dependence on the interaction strength was fitted to the data for not too 
strong interaction. Furthermore, it was noticed that, at strong interaction, 
the persistent current became paramagnetic for all samples, independent of the 
disorder realization.

The Wigner crystal ground state on such a lattice is of square form, as shown 
in Fig.\ \ref{fig:linehop}. Thus, there are two lines with $N_y=2$ in which 
hopping sequences of $L_x=6$ hops can contribute to the persistent current. 
This explains immediately that the response is paramagnetic, since $N_y$ is 
even and $-{\rm sign}(P_S)=+1$. The persistent current decreases in the strong 
interaction limit as $t^6/U^5$. A laborious evaluation of all the 18 sequences 
for each of the lines with $N_y=2$ yields the result
\begin{equation}\label{amp_6x6}
\tilde{I}\rightarrow \tilde{I}^{(6)} \approx 1.808\times 10^6 \frac{t^{6}}{U^5}
\quad\quad \mbox{for} \quad\quad \frac{t}{U}\rightarrow 0 \, .
\end{equation}  
As compared to the previous cases, the prefactor is much larger. This
is caused by the bigger number of contributing sequences and, more importantly,
the difference in interaction energy between the intermediate states
and the unperturbed ground state $\epsilon_\alpha$. The latter can be very
small when the distance between the particles is large. 

A numerical investigation by direct diagonalization of the corresponding 
Hamiltonian matrices \cite{SW_unpub} confirms the signs, the power laws 
and the numerical prefactors predicted by the above formulas, also for the
last case of 4 particles on $6\times 6$ sites, where the persistent
current is indeed found to follow the power law of (\ref{amp_6x6}) at
strong interaction.
However, it must be noticed that the sign of the persistent current is 
well established \cite{benenti} at interaction values
much lower than the ones where the agreement in amplitude with our
formula starts to be good.
Even though the data always follow the power laws at strong interaction,
fitting an exponential interaction dependence, as done in Ref.\ \cite{benenti},
is possible at moderately strong interaction, and might allow to extract useful
informations in the regime where higher order terms of the perturbation theory 
are non-negligible.

\subsection{Disorder effects at strong interaction}
\label{sec:disorder}

The role of the disorder and realization-dependent fluctuations of the 
persistent current vanish in the limit of strong interaction, when 
$W/U \rightarrow 0$. In this section, we treat the lowest order correction in 
$W/U$ to the results presented above.

In addition, we consider the special case of a perfectly clean system $W=0$,
in which the translational symmetry can considerably influence the 
interaction dependence of the persistent current.

\subsubsection{Disorder corrections to the persistent current} 

In order to take into account the disorder corrections in the perturbation 
theory for the longitudinal persistent current, it is not sufficient to
consider in the denominator of the expansion terms the $\epsilon_{\alpha}$ 
which completely neglect the disorder energy. Instead, the full 
energy differences $E_0-E_{\alpha}=-U\tilde{\epsilon}_{\alpha}$ with 
\begin{equation}\label{epsilon_tilde}
\tilde{\epsilon}_\alpha=\frac{1}{U}\left(
E^{\rm I}_{\alpha}+E^{\rm D}_{\alpha}-E^{\rm I}_{0}-E^{\rm D}_{0}\right) 
\end{equation}
account also for the differences in disorder energy between the intermediate 
states $\alpha$ and the ground state. The difference
\begin{equation}
\tilde{\epsilon}_{\alpha}-\epsilon_{\alpha}=\frac{1}{U}
\left( E^{\rm D}_\alpha-E^{\rm D}_0 \right)= 
\frac{W}{U}\sum_{i} v_i \left( n_i(\alpha)-n_i(0) \right) \, ,
\end{equation}
is of the order $W/U$ and vanishes in the limit $W/U \rightarrow 0$. 
With the definition
\begin{equation}
d_\alpha=\sum_{i} v_i \left( n_i(\alpha)-n_i(0) \right)
\end{equation}
we can write $\tilde{\epsilon}_\alpha=\epsilon_\alpha+\frac{W}{U}d_\alpha$
and therewith express the energy difference terms in the
denominator of (\ref{pert_n}) as
\begin{equation}
\frac{1}{E_0-E_{\alpha}} = \frac{-1}{U\tilde{\epsilon}_{\alpha}}
= \frac{-1}
{U\epsilon_\alpha\left(1+\frac{W}{U\epsilon_\alpha}d_\alpha\right)}\, .
\end{equation}

Taking the lowest order term in $W/U$, and averaging over the ensemble yields
\begin{equation}
\left<\frac{1}{E_0-E{\alpha}}\right>\approx \frac{1}
{-U\epsilon_{\alpha}}\left( 1-\frac{W}{U\epsilon_{\alpha}}<d_\alpha>\right)\, ,
\end{equation}
where the brackets $<\dots >$ denote the ensemble average over all disorder 
realizations.
At first glance one could expect that the correction linear in $W/U$
vanishes because $<d_\alpha>=0$, when the disorder average is taken over all 
values of $d_{\alpha}$. However, since the ground state is 
the Wigner crystal pinned at the {\em lowest} disorder configuration, 
$d_\alpha$ is more likely positive than negative. The first correction 
is thus linear in $W$, and since $\epsilon_\alpha$ is always positive, the 
lowest order correction to the persistent current due to disorder is
reducing the contributions of all sequences, the result being
\begin{eqnarray}\label{disorder_correction}
I^{(L_x)}_{\rm W} &\approx& -2 \frac{t^{L_x}}{U^{L_x-1}} \sum_{S_{\rm f}}
\frac{{\rm sign}(P_S)}
{\epsilon_{\alpha_1}\epsilon_{\alpha_2}\dots
  \epsilon_{\alpha_{L_x-1}}}
\nonumber\\
&\times &\left(
 1-\sum_\alpha\frac{1}{\epsilon_{\alpha}}\frac{W<d_\alpha>}{U}
+O[(W/U)^2]\right)\, .
\end{eqnarray}

\subsubsection{The particular case $W=0$}

At $W=0$ we have $H_0=H_I$, and the ground state at $t=0$ is degenerate 
because of the translation symmetry. The number $n_{\rm D}$ of equivalent 
degenerate Wigner crystal positions and corresponding basis states 
$|\psi_0^{(\beta)}\rangle $ depends on the system size and the number of
particles. The hopping terms however lead to a coupling of these
degenerate basis states, and split the degenerate levels, except
for special situations where symmetries persist. The coupling terms
themselves can be expressed in terms of a perturbative expansion in
$t$. The order $p$ in which the degeneracy of the ground state is
lifted may be different from the lowest order $n$ which yields a
finite persistent current. We address in the following the three
different cases which may occur.

\paragraph{$p>n$}  

In this case, the splitting can be ignored as
compared to the persistent current at strong interaction. The perturbation
theory can be applied as in the disordered case, using one of the
degenerate ground state configurations (the result will be the same
for arbitrary superpositions of the degenerate basis states). 

\paragraph{$p=n$} 

When the persistent current is given by terms which are of the same 
order as the ones which lift the ground state degeneracy, the flux
dependent correction to the ground state energy and the persistent
current are given by the lowest eigenvalue of the effective coupling
matrix $M$ between the $n_{\rm D}$ degenerate basis states. The
matrix elements are 
\begin{eqnarray}\label{eff_matrix}
&&M_{\beta,\beta'}=\sum_{\alpha_1,\alpha_2,\dots ,\alpha_{p-1}} \\
&&\frac{
\langle \psi_0^{(\beta)} | H_{\rm K} | \psi_{\alpha_1} \rangle 
\langle \psi_{\alpha_1} | H_{\rm K} | \psi_{\alpha_2} \rangle 
 \dots 
\langle \psi_{\alpha_{p-1}} | H_{\rm K} | \psi_0^{(\beta')} \rangle }
{(E_0-E_{\alpha_1})(E_0-E_{\alpha_2})\dots (E_0-E_{\alpha_{p-1}})} \, . 
\nonumber
\end{eqnarray} 
Since in this case all matrix elements are $\propto t^n/U^{n-1}$, the
lowest eigenvalue and thus the persistent current will follow such a
power law too.  

\paragraph{$p<n$}

When the levels are split at an order which is lower than the one
which yields contributions to the persistent current, the higher order
terms must be calculated using the lowest energy ground state found
from the diagonalization of $M$. Such a ground state will be a
superposition 
\begin{equation}
|\psi_0 \rangle=\sum_{\beta=1}^{n_{\rm D}}f_\beta|\psi_0^{(\beta)}\rangle
\quad\quad \mbox{with} \quad\quad \sum_\beta^{n_{\rm D}} |f_\beta|^2 = 1
\end{equation}
of the different Wigner crystal positions.
Plugging this ground state into the general expression for the corrections to 
the ground state energy (\ref{pert_n}), one gets
\begin{eqnarray}\label{no_disorder}
&&E^{(n)}_0(W=0)=\sum_{\beta,\beta'}\; f_\beta^* f_{\beta'}^{\phantom{*}}
\sum_{\alpha_1,\alpha_2,\dots ,\alpha_{n-1}} \\
&&\frac{
\langle \psi_0^{(\beta)} | H_{\rm K} | \psi_{\alpha_1} \rangle 
\langle \psi_{\alpha_1} | H_{\rm K} | \psi_{\alpha_2} \rangle 
 \dots 
\langle \psi_{\alpha_{n-1}} | H_{\rm K} | \psi_0^{(\beta')} \rangle }
{(E_0-E_{\alpha_1})(E_0-E_{\alpha_2})\dots (E_0-E_{\alpha_{n-1}})} \, , 
\nonumber
\end{eqnarray} 
and realizes that contributions can arise from hopping processes starting at
any of the ground state components $|\psi_0^{(\beta)}\rangle$, and
ending at an arbitrary $|\psi_0^{(\beta')}\rangle$ where $\beta$ and
$\beta'$ can be equal or different.  
The processes with $\beta\neq \beta'$ can in some special cases have a
lower power in $\frac{1}{U}$ than the processes with $\beta=\beta'$,
relevant in the disordered case, and therefore dominate in the limit
of strong interaction. 

The case $W=0$ can be qualitatively different from the disordered case
$W\neq 0$, even at very weak disorder. The reason for this is the fact
that the unperturbed ground state is a superposition of the degenerate
positions of the Wigner crystal. This degeneracy is lifted by the
disorder, and the typical energy difference 
between two positions is of the order $W\sqrt{N}$ (this arises from the $N$ 
different random on-site energies, all being of order $W$). The coupling 
between these electro-statically equivalent states is provided by the hopping 
terms and can itself be estimated from a perturbation theory. The lowest
order coupling needs a number of hops which is typically $p=N$ since the
Wigner crystal must in most cases be translated as a whole and each of
the particles must hop at least once. In some cases, a $p<N$ can be
sufficient, as in the example of 3 particles on $4\times 2$ sites 
(see below).

Then, the coupling decreases at strong interaction as $t^p/U^{p-1}$ and
becomes, in the limit of strong interaction $U$, 
always much smaller than the splitting due to the disorder. This holds
at any finite disorder value (provided $p>1$), and prevents a mixing of
the different Wigner crystal configurations in the ground state. Only
for $W=0$, the splitting by the disorder vanishes, and an
infinitesimal coupling leads to the superposition of the different 
configurations.

\subsubsection{Example for disorder corrections}

\paragraph{$4\times 2$ sites, 4 particles} 

The disorder dependence of the flux sensitivity is illustrated for the example 
of 4 particles in $(L_x,L_y)=(4,2)$. One expects an $1/U^3$ dependence for the 
leading term of the persistent current (\ref{amp_4x2}) and, according to 
(\ref{disorder_correction}), a $W/U^4$ dependence for the linear correction in 
disorder. We have performed numerical calculations at strong interaction
\cite{SW_unpub}, which are in good agreement with these power laws
predicted by the lowest order of the perturbation theory. The persistent 
current decreases when the strength of the disorder is increased, as
predicted by the negative correction in formula (\ref{disorder_correction}). 

\paragraph{Effects of degeneracies at $W=0$}

The clean case however has a very different 
amplitude due to the degeneracies in the Wigner crystal. 
The persistent current can be calculated from (\ref{no_disorder}), and after
considering all the possible sequences between the different
degenerate configurations, one obtains
$\tilde{I}^{(4)}(W=0)\approx 2.1238\, \tilde{I}^{(4)}$, with
$\tilde{I}^{(4)}$ taken from (\ref{amp_4x2}). A numerical check \cite{SW_unpub}
confirms this result at strong interaction. 

\paragraph{$4\times 2$ sites, 3 particles, $W=0$}
\begin{figure}[tb]
\centerline{\epsfxsize=4cm\epsffile{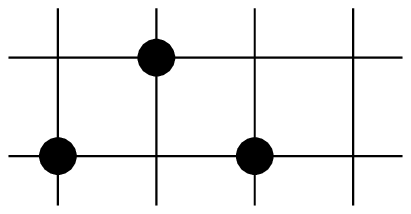}
\hfill \epsfxsize=4cm\epsffile{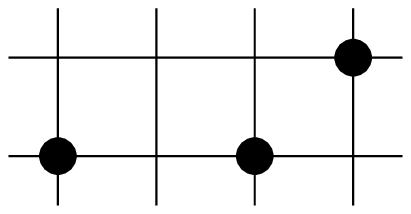}}
\vspace{3mm}
\caption[fign=3]{\label{fign=3} Two degenerate configurations for 3 
particles on $4\times 2$ sites which can be linked to each other by
sequences of only two hopping processes.}
\end{figure}
In order to show that the absence of disorder does not only
change the prefactor of the interaction dependence at large $U$, but can even 
influence the power of the $U$-dependence, we consider 3 particles in 
$(L_x,L_y)=(4,2)$. The particle which is alone in one line has two possible 
positions which are degenerate at $t=W=0$, and there exists a sequence of
only two one-particle hops which links the two degenerate states with each 
other (see Fig.\ \ref{fign=3}). Since the ground state at $W=0$ is a 
superposition of all possible degenerate configurations, this process leads 
to a contribution $\propto t^2/U$ to the persistent current which 
dominates at strong interaction. 
In the disordered case, in contrast, the degeneracy of the ground state 
configurations is lifted and only sequences consisting of at least
four hops can link the ground state to itself. Then, the interaction 
dependence of the lowest order contribution to the persistent current
is $\propto t^4/(W U^{2})$. Interestingly, in this example, one of the
intermediate states touched by a sequence of 4 hops can be a different, but 
electro-statically equivalent configuration of the Wigner crystal, and 
$\epsilon_\alpha=0$, such that even at $U\gg W$, the disorder energy cannot be 
neglected.

A numerical check \cite{SW_unpub} shows that by adding the disorder, 
the asymptotic dependence indeed follows this prediction and goes 
from $\propto U^{-1}$ to $\propto U^{-2}$.     
 
\subsection{Transverse current}
\label{sec:transverse}

\subsubsection{General considerations}

Without disorder, the current in $y$-direction must vanish because of the 
symmetry of the problem (no flux is applied in $y$-direction). 
While this remains true in the ensemble average, the presence of disorder 
breaks this symmetry in individual samples. 
A finite transverse current can be observed in addition to the longitudinal 
current driven by the flux $\phi_x$.
The transverse current can be calculated from
\begin{equation}
I_t(\phi_x)=-\left. \frac{\partial E_0(\phi_x,\phi_y)}
{\partial \phi_y}\right|_{\phi_y=0} \, ,
\end{equation}
such that a theoretical approach to determine $I_t(\phi_x)$ consists in using 
the perturbative expansion for the ground state energy, as for the longitudinal
current, but with a finite $\phi_y$. Then, the derivative with respect to
$\phi_y$ at $\phi_y=0$ can be evaluated for the leading contribution in the 
strong interaction regime. Therefore, only terms in the series (\ref{pert_n}) 
which depend on $\phi_y$ can make a finite contribution to the 
transverse current. 

In order to get such contributions, one needs hopping sequences in 
$y$-direction which contain hops crossing the boundary between sites 
$(x,L_y)$ and $(x,1)$ with $h_{\rm u}-h_{\rm d}\neq 0$. $h_{\rm u}$ and 
$h_{\rm d}$ denote the number of such hops from $y=L_y$ to $y=1$ ('upwards')
and from $y=1$ to $y=L_y$ ('downwards').
 
However, the flux dependence of the energy due to sequences containing $L_y$ 
upward or $L_y$ downward hoppings is $\propto \cos\, \phi_y$, as for the 
longitudinal processes (\ref{energy_lowest}), which are 
$\propto \cos\, \phi_x$. As a consequence, the resulting transverse 
current from these sequences is $\propto \sin\, \phi_y$, and vanishes when 
$\phi_y=0$ is taken, just like the longitudinal current (\ref{pc_lowest})
at $\phi_x=0$. Therefore, no transverse current can be obtained in the 
order $L_y$ of the perturbation theory.

\begin{figure}[tb]
\centerline{\epsfxsize=2in\epsffile{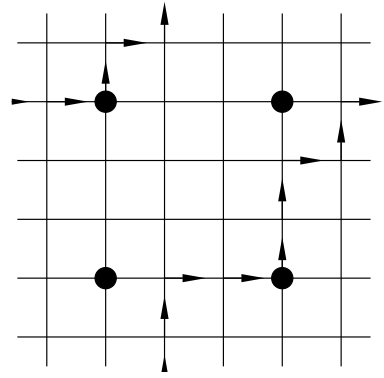}}
\caption[linehop]{\label{fig:diaghop} A sequence of hopping processes which 
gives rise to a finite contribution to the transverse current in the 
disordered system of 4 particles on $6\times 6$ sites. The arrows indicate 
$L_x+L_y=12$ forward and upward hops.}
\end{figure}
Nevertheless, higher order sequences exist which do lead to finite 
contributions. These are due to sequences of hops which cross both, the 
boundary in $x$, and the boundary in $y$-direction. The lowest order of the 
expansion in which transverse currents appear is $m=L_x+L_y$, corresponding to 
sequences which consist of a combination of only forward and downward hops,
only backward and upward hops, or vice versa, which complete one round
in both spatial dimensions of the system. An example of such a sequence 
is shown in Fig.\ \ref{fig:diaghop}.
 
As before, each of the sequences containing only forward and upward
hops  $S_{\rm fu}$ can be associated with the reverse process consisting of  
backward and downward hops $S_{\rm bd}$. The contribution of the
reverse process differs only in the sign of the flux-dependent phase
factors. The correction to the ground state energy (\ref{pert_n}) of
all  $S_{\rm fu}$ and $S_{\rm bd}$ sequences together can be written as
\begin{equation}\label{exy_lowest_fu}
E_{0,{\rm fu+bd}}^{(m)} = -2 \frac{t^{m}}{U^{m-1}} \cos(\phi_x+\phi_y) 
\sum_{S_{\rm fu}} \frac{{\rm sign}(P_{S_{\rm fu}})}
 {\tilde{\epsilon}_{\alpha_1}\dots \tilde{\epsilon}_{\alpha_{m-1}}} \, .
\end{equation} 
The same association can be made for sequences containing only forward and 
downward hops $S_{\rm fd}$, which are regrouped with the corresponding reverse 
processes consisting of backward and upward hops, yielding
\begin{equation}\label{exy_lowest_fd}
E_{0,{\rm fd+bu}}^{(m)} = -2 \frac{t^{m}}{U^{m-1}} \cos(\phi_x-\phi_y) 
\sum_{S_{\rm fd}} \frac{{\rm sign}(P_{S_{\rm fd}})}
 {\tilde{\epsilon}_{\alpha_1}\dots \tilde{\epsilon}_{\alpha_{m-1}}} 
\, .
\end{equation} 

\subsubsection{Lowest order result for the transverse current}

By taking the derivative with respect to $\phi_y$ at $\phi_y=0$, we obtain for 
the transverse current in $m$-th order of the perturbation theory
\begin{eqnarray}\label{pct_lowest}
&&I_t^{(m)}(\phi_x) = 2 \frac{t^{m}}{U^{m-1}} \sin\phi_x \\ \nonumber 
&&\times \left(
\sum_{S_{\rm fu}} \frac{{\rm sign}(P_{S_{\rm fu}})}
 {\tilde{\epsilon}_{\alpha_1}\dots \tilde{\epsilon}_{\alpha_{m-1}}} 
-\sum_{S_{\rm fd}} \frac{{\rm sign}(P_{S_{\rm fd}})}
 {\tilde{\epsilon}_{\alpha_1}\dots \tilde{\epsilon}_{\alpha_{m-1}}} 
\right)\, ,
\end{eqnarray} 
with $m=L_x+L_y$ being the lowest order which yields a non-zero contribution.
Even when $U\gg W$, the disorder energy cannot be neglected as in the lowest 
order term for the longitudinal current, since the two contributions to the 
current in (\ref{pct_lowest}) cancel each other exactly when the disorder 
vanishes. It can also be seen that the lowest order result for the transverse
current is proportional to $\sin\phi_x$ and vanishes with the longitudinal 
current at $\phi_x=0$. 

\subsubsection{Example}

In the case of 4 particles in a lattice of $6\times 6$ sites, we have seen 
above (see equation (\ref{amp_6x6})) that the longitudinal persistent current 
decreases at strong interaction following the power law $\propto t^6/U^5$. The 
suppression of the transverse current is much more pronounced, since the 
leading contribution at strong interaction is of the order $L_x+L_y=12$, and 
our theory (\ref{pct_lowest}) predicts that it decays $\propto t^{12}/U^{11}$. 
This means that the particle mobility is more and more restricted to the 
longitudinal direction when the interaction is increased. That the suppression 
of the transverse current is much stronger than the one of the longitudinal 
persistent current has been noticed in the numerical study of 
Ref.\ \cite{benenti}. The dominance of the longitudinal current at
strong interaction also explains the observation of ``plastic flow''
in a study of the local persistent currents \cite{berko2}. 
The orientation of the local currents has been proposed as a signature
of the insulator-metal transition occurring in interacting 2d systems 
\cite{benenti_mor}.
 
\subsection{Persistent current for electrons with spin}
\label{sec:spin}

In this section we present some examples of the effects of the
spin of the electrons on the longitudinal persistent current, showing
that our approach is not restricted to 
spinless fermions, but can also be used for electrons carrying spin.

For finite systems containing $N$ particles, it is well known that the spin 
polarization of the ground state can depend on the magnetic flux due to level 
crossings between states of different spin symmetry as a function of
the flux \cite{haeusler}. 
However, without an external magnetic field acting on the electrons in the 
ring, there is a degeneracy related to the operator $S_z$ in the subspace of 
fixed total spin and one can write $E(S^2,S_z)=E(S^2)$. 

It is in principle possible (though difficult experimentally) to create a 
magnetic flux through the ring while the magnetic field remains vanishing at 
the positions of the electrons in the ring. Such a flux does not lift the spin 
degeneracy mentioned above. If we now want to follow the dependence of the 
ground state energy on the flux, we can restrict the study to the subspace of 
minimum absolute value of $S_z$, choosing $S_z=0$ for an even number of 
particles with $N_\uparrow=N/2$ spins up and $N_\downarrow=N/2$ spins down, 
or $S_z=1/2$ for an odd number of particles with $N_\uparrow=\frac{N+1}{2}$ 
spins up and $N_\downarrow=\frac{N-1}{2}$ spins down.

\subsubsection{One-dimensional systems}

In 1d, a general rule exists for the sign of the persistent current in the case
of spinless fermions \cite{leggett}. This rule is valid at arbitrary disorder 
and interaction. Below, we address the question of the sign of the
persistent current for electrons with spin in different parameter regimes. 

\paragraph{Non-interacting electrons with spin}

In the case of non-interacting electrons with spin, one can separately consider
the flux dependence of the energy of the particles with up spin and the one of 
the particles with down spin, and add up their contributions to the persistent 
current. 

If $N$ is even, the contribution of the $N/2$ up-spins to the persistent 
current will have the same sign $(-1)^{N/2}$ as the contribution of the $N/2$ 
down-spins. This determines the sign of the persistent current. For odd $N$, 
the contribution of the $(N+1)/2$ up spins will have the sign $(-1)^{(N+1)/2}$ 
while the down spins contribute a term with sign $(-1)^{(N-1)/2}$. In this 
case, to know the sign of the persistent current one must compare the 
amplitudes of the two contributions and the result will in general
depend on the disorder configuration. However, in the zero-disorder
case at low filling, the sign of the persistent current around flux 
$\phi=0$ is known to be paramagnetic for odd $N>1$ \cite{loss-gold}.

\paragraph{$N$ even, strong interaction}

At strong interaction, the charges form a Wigner crystal which is pinned by 
the disorder. In this limit, the spin degree of freedom can be treated by an 
effective spin Hamiltonian. The latter turns out to be the anti-ferromagnetic 
Heisenberg Hamiltonian when $\phi=0$, the positions  $i={i_1,i_2,\dots,i_N}$ of
the charges of the Wigner crystal being the spin lattice sites. Thus, the 
expectation values of the occupation number operators vanish everywhere except 
on these sites where they satisfy 
$<\hat{n}_{i_k,\uparrow}+\hat{n}_{i_k,\downarrow}>=1$. 
For an even number $N$ of spins in 1d, according to a theorem by Marshall, the
ground state $|\psi_0\rangle$ of this spin Hamiltonian is a singlet of total 
spin $S=0$ \cite{auerbach}, and can be expressed in the form 
\begin{equation}\label{spin_gs}
|\psi_0\rangle =\sum_\beta f_\beta 
\prod_{i=i_1}^{i_N}\left( c^{+}_{i,\uparrow }\right)^{n_{i,\uparrow}(\beta)}
\prod_{i=i_1}^{i_N}
\left( c^{+}_{i,\downarrow }\right)^{n_{i,\downarrow}(\beta)} |0\rangle
\end{equation}
with real $f_\beta > 0$ for all spin configurations $\beta$ with fixed
$S_z=0$. Note that in this expression, the ordering of the operators is done
firstly according to spin, and secondly according to the position. 

As in the case of spinless fermions without disorder, the ground state is a 
superposition of different basis states and the lowest order contributions to 
the ground state energy, which are flux dependent, are again of the order 
$n=L_x$. Similar to (\ref{no_disorder}), they can arise from different kinds 
of sequences. First, only the up spins are moved around the ring, giving rise 
to a cyclic permutation of the spin up operators in (\ref{spin_gs}), yielding 
the sign $(-1)^{N_\uparrow}$. The sequences involving only down spins give the 
same result. In addition, a sequence which moves all of the particles to the 
position of their neighbor can also contribute to the flux dependence since 
the resulting spin configuration is also contained in the ground state. The 
sign however is the same as the one of the previous sequences, since only one 
of the particles crosses the boundary and therefore only the order of the 
operators for one spin direction has to be restored by the corresponding cyclic
permutation, and because the prefactor $f_{\beta}^* f_{\beta'}^{\phantom{*}}$,
arising from the weights of the different components in the ground state, is 
always real and positive. 

Since all contributions have the same sign, the sign of the persistent current 
in one-dimensional disordered chains at strong interaction is always given by 
this sign $(-1)^{N/2}$, provided the particle number is even,
independent of the particle density $N/L$. 
Because the ground state (\ref{spin_gs}) holds only at $\phi=0$, this
sign rule is granted only in the vicinity of $\phi=0$. 

The resulting sign is the same as the one found in the non-interacting 
case \cite{loss-gold}, and for the Hubbard model at half filling
\cite{fye}, but it differs from the result
for the Hubbard model at low filling \cite{yu,kotlyar}, where
the current around $\phi=0$ is found to be diamagnetic for even
numbers of particles at strong interaction. 
This difference may be due to the fact that the Hubbard interaction is
local, making the Hubbard model at low filling in the $U\rightarrow \infty$ 
limit equivalent to spinless fermions. The more realistic long-range Coulomb
interactions considered here do not show this artifact and lead to a 
qualitatively different behavior. The result is also different from
the one obtained for Coulomb interacting electrons with spin in
a continuous model with one barrier \cite{haeusler}, which is always 
found to be diamagnetic at strong interaction. In this case, one may
attribute the result to the fact that the interaction could lead to a
rigid Wigner crystal which, in a continuous model, would be equivalent to one
single heavy particle. 

\subsubsection{Two-dimensional systems}

For two dimensional systems, there is the same degeneracy in the position of 
the up and the down spins in the minimum $S_z$ subspace. In the strong 
interaction limit, each line of the Wigner crystal should follow the one 
dimensional law explained above, but the different spin configurations
contributing to the ground state in 2d may have different numbers of
spin up and spin down electrons in a given line. This yields
contributions of different signs to the persistent current which have
to be considered explicitly. It seems not to be possible to provide a
simple sign rule. However, the power law decay of the persistent
current at large interaction strength should have the same exponent as
in the spinless case.

\subsection{Size dependence and limitation of the theory}
\label{sec:size}

Here, we address the size dependence of the persistent current
and the limitation of our theory which is given by the presence of defects
in the Wigner crystal. 

\subsubsection{Localization length for the Wigner crystal}

From the exponential size-dependence of equation (\ref{amplitude_lowest}),
one can extract the localization length for the Wigner crystal. A 
comparison to the exponential size scaling
\begin{equation}
I(L_x)\propto \exp(-L_x/\xi)
\end{equation}
yields for large $L_x$ a localization length
\begin{equation}
\xi=\left(\ln \left(\frac{U}{t}\right)\right)^{-1}\, ,
\end{equation}
showing the same decrease with increasing interaction strength as the 
localization length of the Mott insulator appearing in the Hubbard
model at strong interaction \cite{stafford}. 

\subsubsection{Defects in the Wigner crystal}

In the thermodynamic limit, the Wigner crystal is no longer a single
domain and the above theory cannot be applied in the present form.
At any finite, even very strong interaction, defects and domain walls can 
arise which allow to gain an amount of potential energy which grows with the 
size of the domains. 
The crystal might then prefer to be divided in defect-free domains
each being pinned by the disorder. In the weak disorder limit, one can
use similar arguments as in \cite{Imry} in order to estimate the size
of such domains. In the following, we briefly address different types of 
defects and calculate the associated critical ratio
 $\left( \frac{U}{W}\right)_c$ above which the theory applies.

\paragraph{Point defects}
In the case of point-like defects, consisting of a single charge of
the Wigner crystal being deplaced by one lattice constant $a$ with
respect to its position in the perfect crystal, the gain in disorder 
potential energy can be estimated to be 
$\Delta E^{\rm point}_{\rm disorder}\approx W$.
On the other hand, the cost in interaction energy is 
$\Delta E^{\rm point}_{\rm interaction}\approx 
\frac{Ua^2}{b^3}$ with $b$ being the
mean spacing between particles. Since the charge density is given by 
$\nu=1/b^2$, this yields the density depending criterion
\begin{equation}
\frac{U}{W} >\left( \frac{U}{W}\right)^{\rm point}_c\approx \frac{1}{a^2 \nu^{3/2}}
\end{equation}
for the stability of the Wigner crystal against the creation of point
defects.

\paragraph{Domain walls}
The creation of a domain wall costs an interaction energy which is
of the order $\Delta E^{\rm wall}_{\rm interaction}\approx 
\frac{Ua^2}{b^3} L^{\rm wall}$ with $L^{\rm wall}$ being the length of
the wall. By deplacing a domain of linear size $R$ containing
$N_d=R^2\nu$ particles, one can
typically gain an amount $W\sqrt{N_d}$ of disorder potential
energy. However, the first domain-like defect will appear in the most 
favorable position. Since the maximally possible gain is the extremely
unprobable value $W N_d$, one can speculate that the gain in the
optimal position is given by an intermediate power  
$\Delta E^{\rm domain}_{\rm disorder}\approx W N_d^\gamma$ with 
$1/2 \le \gamma < 1$.
The length of the corresponding wall around the domain is 
$L^{\rm wall}\sim \sqrt{N_d}$ such that this results in the stability
of the crystal against the creation of domains of size $R$ for
\begin{equation}
\frac{U}{W} > \left( \frac{U}{W}\right)^{\rm domain}_c (R)
\approx \frac{R^{2\gamma-1}}{a^2 \nu^{2-\gamma}}\, .
\end{equation}
At a fixed W, if $U>U^{\rm domain}_c(L)$, a crystal of size $L$ will
not be perturbed by domains. Otherwise, one can use  
$U=U^{\rm domain}_c(R_c)$ to extract the typical domain size $R_c$.

In systems larger than $R_c$, the electron crystal is divided in 
many domains of size $R_{c}$. If we neglect the coupling energy between these 
domains, a rough estimate for the response to a flux $\phi_x$ in 
$x$-direction of a system of size $(L_x,L_y)$ is to take the
product of the amplitudes of the longitudinally aligned $L_x/R_c$ domains and 
to sum over the responses of the $L_y/R_c$ ``channels''.

Although we have not addressed all possible kinds of defects, we can 
assume that there is always a critical threshold 
$\left( \frac{U}{W}\right)^{\rm defect}_c$ above which our perturbative
theory applies.
 
\section{Summary}

We have presented a systematic perturbative treatment of persistent
currents in 2d lattice models for the case of strong Coulomb interaction, when 
the electronic charge density forms a Wigner crystal. The contribution with the
weakest interaction dependence corresponds to sequences of one-particle hops 
along the shortest paths around the ring. These sequences dominate in the 
limit of strong interaction and determine the sign of the persistent current.
For spinless fermions, this sign follows simple rules which depend only on the 
structure of the Wigner crystal.

Furthermore, we have shown that, except special cases, the leading term for 
the persistent current and therewith the sign of the persistent current at 
strong interaction do not depend on the realization of the disordered 
potential. Only the complete absence of disorder can qualitatively change the 
behavior. We considered the disorder corrections systematically and showed 
that they decay as $W/U$ at strong interaction.

In addition, we have shown that transverse currents appearing in individual 
disordered samples are suppressed much faster than the longitudinal current, 
thereby establishing an orientation of the local currents in longitudinal 
direction when the interaction is increased. 
This explains the numerical observation of the realization-independent sign
of the persistent current and of the orientation of the local currents in 
longitudinal direction reported in 
Refs.\ \cite{benenti,benenti_mor,berko1,berko2}. 

\begin{acknowledgement}
We thank G.\ Benenti, W.\ H\"ausler, G.-L.\ Ingold, D.\ Loss, J.-L.\ Pichard, 
P.\ Schwab, and X.\ Waintal for useful discussions. Particularly
helpful discussions with R.\ Jalabert are gratefully acknowledged.
Partial support from the TMR network ``Phase coherent dynamics of hybrid 
nanostructures'' of the EU and from the Procope program of the 
DAAD/A.P.A.P.E. is gratefully acknowledged.
\end{acknowledgement}

\end{document}